# Structural Brain Connectivity and Treatment Improvement in Mood Disorder

Sébastien Dam[1], Jean-Marie Batail[2,3], Gabriel H. Robert[1,2,3], Dominique Drapier[2,3], Pierre Maurel[1] & Julie Coloigner[1]

[1] Univ Rennes, Inria, CNRS, IRISA, INSERM, Empenn U1228 ERL, F-35000 Rennes, France

[2] Centre Hospitalier Guillaume Régnier, Academic Psychiatry Department, F-35703 Rennes, France

[3] CIC 1414, CHU de Rennes, Inserm, Rennes, France

Corresponding author: Julie Coloigner, julie.coloigner@irisa.fr

**Abstract**

***Background***: The treatment of depressive episodes is well established, with clearly demonstrated effectiveness of antidepressants and psychotherapies. However, more than one-third of depressed patients do not respond to treatment. Identifying the brain structural basis of treatment-resistant depression could prevent useless pharmacological prescriptions, adverse events, and lost therapeutic opportunities.

***Methods***: Using diffusion magnetic resonance imaging, we performed structural connectivity analyses on a cohort of 154 patients with mood disorder (MD) – and 77 sex- and age-matched healthy control (HC) participants. To assess illness improvement, the MD patients went through two clinical interviews at baseline and at 6-month follow-up and were classified based on the Clinical Global Impression-Improvement score into improved or not-improved. First, the threshold-free network-based statistics was conducted to measure the differences in regional network architecture. Second, nonparametric permutations tests were performed on topological metrics based on graph theory to examine differences in connectome organization.

***Results***: The threshold-free network-based statistics revealed impaired connections involving regions of the basal ganglia in MD patients compared to HC. Significant increase of local efficiency and clustering coefficient was found in the lingual gyrus, insula and amygdala in the MD group. Compared with the not-improved, the improved displayed significantly reduced network integration and segregation, predominately in the default-mode regions, including the precuneus, middle temporal lobe and rostral anterior cingulate.

***Conclusions***: This study highlights the involvement of regions belonging to the basal ganglia, the fronto-limbic network and the default mode network, leading to a better understanding of MD disease and its unfavorable outcome.

**Impact Statement**

Different treatments have been developed for patients suffering from mood disorder (MD). Yet, more than one-third of depressed patients do not respond to them. The present study provides statistical analysis of brain structural networks using the threshold-free network-based statistics and graph theoretical approaches based on a large cohort of MD patients. This enabled the identification of salient regions involved in MD and the process of treatment-resistant depression. Some of the findings are coherent with the results based on functional connectivity, which strengthens the motivation to jointly study both structural and functional connectivity.

1  **Introduction**

Mood disorder (MD) is a mental health condition, with occurrences of depressive episodes that impact nearly all aspects of a person's daily life (Corponi et al., 2020). Despite the improvement and increased availability of treatments since the 1980s, no decrease of depressive episodes or prevalence have been reported (Ormel et al., 2022). The emergence of the COVID-19 pandemic even contributed to a 25% increase in the global prevalence of depression in 2020 according to the World Health Organization. Previous studies have shown that only 40 to 60% of Major Depressive Disorder (MDD) patients respond to their first-line treatment (Carvalho et al., 2007; Gartlehner et al., 2011; Trivedi et al., 2006) and that the remission rate with standard antidepressant treatments reaches 30-40% (Rush et al., 2006). Lack of appropriate treatment could lead to increased duration of symptoms, greater risk of suicide and trial-and-error pharmacological prescriptions strategy, which might be responsible for adverse events and induce increased chronicity. It is therefore crucial to identify markers that would help identify individuals at risk of pharmacological resistance and opt for alternative therapeutic strategies.

Over the past twenty years, magnetic resonance imaging (MRI) has played an important role in deciphering the pathophysiology of MDD (Zhuo et al., 2019). Some MDD studies found widespread gray matter changes in the frontal lobe and the hippocampus (Arnone et al., 2012; Zou et al., 2010), as well as volume reductions in prefrontal and anterior cingulate cortices, and also in the basal ganglia (caudate, putamen, globus pallidus) (Bora et al., 2012). Other findings highlighted key regions involved in emotional processing like the amygdala (Sandu et al., 2017), insula (Johnston et al., 2015) or thalamus (Batail et al., 2020) in treatment-resistant depression (TRD). Besides volumetric MRI analysis, other works have investigated the brain structural connectivity in MDD, especially with the emergence of advanced neuroimaging technics such as diffusion-weighted imaging (DWI). Numerous DWI studies revealed significant white matter modifications, such as decreased fractional anisotropy (FA) in the corpus callosum, bilateral anterior limb of the internal capsule and the superior longitudinal fasciculus (Ota et al., 2015), suggesting a dysfunction in cortical-subcortical pathways (Coloigner et al., 2019; van Velzen et al., 2020). Previous diffusion tensor imaging (DTI) studies also reported decreased white matter integrity in subcortical regions including the thalamus, caudate, and putamen in patients with late-life depression (LLD) (Alexopoulos et al., 2009; Yuan et al., 2007). More recently, a review focusing on functional and structural connectivity in MDD patients under treatment found modified FA in the white matter tracts of multiple regions, such as the hippocampus, raphe nucleus or amygdala in treatment responders compared to non-responders. Alterations in other DTI parameters, namely the axial diffusivity and radial diffusivity were also associated with clinical response in white matter tracts (Tura and Goya-Maldonado, 2023). In these articles, voxel-wise combined with mass-univariate linear analyses were key in identifying these regions. Here, we propose to extend these methods by performing a connectomics analysis (Hagmann et al., 2007) in order to highlight the disrupted network connections in depression.

Recently, graph theoretical approaches have been increasingly employed to study cerebral networks. This framework usually models the brain as a complex graph composed of nodes associated with regions of interest (ROIs) and connections representing the density of white matter tracts connecting cortical and subcortical regional nodes (Bullmore and Bassett, 2011). By mapping the whole brain as an interconnected network, one can conduct statistical analyses on both connections and nodes to unravel the human connectome. On the one hand, the network-based statistic (NBS) has been widely used in the last decade to examine variations on connections, as it controls the family-wise error (FWE) by hypothesizing that altered edges form connected components (Zalesky et al., 2010). However, its statistical properties have been questioned due to the necessity of relying on an arbitrary threshold in the first step of the algorithm. More recently, the threshold-free cluster network-based statistics (TFNBS) based on the NBS was designed to remove this prior parameter by using multiple thresholding steps, while displaying statistical power at least similar to that of the NBS (Baggio et al., 2018). On the other hand, graph theory is also increasingly considered to perform analyses on nodes as it can be used to compute topological metrics. They quantify aspects of integration and segregation in order to detect differences in connectome organization (Rubinov and Sporns, 2010).

With these advances in graph theoretical approaches, several studies employed such innovative yet validated connectomics technics described above in order to find biomarkers of MDD and TRD (Korgaonkar et al., 2020, 2014; Yun and Kim, 2021). A recent study assessed network differences in functional connectivity to identify connectomic predictors of treatment response using the NBS (Korgaonkar et al., 2020). In this work, a connectomic signature characterized by an elevated intra-network intrinsic functional connectivity was identified within the default mode, fronto-parietal and somatomotor brain networks. Using graph theory, another functional study reported an increase in degree centrality (number of links held by each node) and local efficiency (measure of local network connectivity) in regions mostly belonging to the default mode network (DMN) in MDD patients compared to healthy controls (HC) (Ye et al., 2015). In functional brain networks, high degree centrality suggests a topological reorganization of nodes around central hubs while high local efficiency is indicative of segregated neural processing (Rubinov and Sporns, 2010). Graph metrics also displayed significant differences in structural connectivity as illustrated by higher nodal efficiency in the thalamus, inferior-middle-superior temporal gyri and lingual gyrus in MDD patients compared to HC (Yun and Kim, 2021; Zheng et al., 2019), but lower nodal efficiency in the thalamus, putamen and caudate nucleus in remitted LLD patients compared to HC (Wang et al., 2020) The changes in regions of the basal ganglia, either in volumetric, DWI or structural connectome studies, highlight the role that these structures play in the patterns of depressive symptoms (Lafer et al., 1997). In fact, the basal ganglia have long been studied in patients with mood disorder mainly because of its role in motor response, response to reward, and in motivation (Macpherson and Hikida, 2019). Thus, its investigation remains a key concern with regard to the pathogenesis and the clinical management of depressive disorders, especially in a context with more advanced technics in brain imaging.

Yet, despite all these findings, some studies revealed that graph metrics were not significantly different between structural connectomes of MDD patients and HC (Korgaonkar et al., 2014; Liu et al., 2018; Xu et al., 2021), or within the MDD patients with different disease status (acute, partial remission, full remission) (Repple et al., 2020). This suggests that there is no strict consensus on the brain regions involved in MD and studying whole-brain structural connectivity is thus necessary to capture patterns of MD.

In this paper, based on the literature, we hypothesized that the brain regions belonging to the basal ganglia and the fronto-limbic network would likely characterize MD patients versus HC, while those from the DMN would be a marker of treatment resistance at 6-month follow-up in our sample. To assess the network modifications, we conducted statistical connection-wise and node-wise analyses to estimate structural patterns of MD and its unfavorable outcome. We performed the TFNBS on structural graphs and permutation tests on topological metrics, based on a large cohort of MD patients compared to HC. Some of the patients were followed over 6 months to study illness improvement and were thus defined as improved (I) or not-improved (NI). Therefore, we also examined whole-brain connectomics between the I and NI groups with these methods.

## 2 Materials and Methods

### 2.1 Participants and study design

This prospective cross-sectional cohort study included 154 MD patients. They were recruited from routine care units in the psychiatric university hospital of Rennes between November 2014 and January 2017. Prior to enrollment, all patients were informed of the study complete description, and their written informed consent was obtained. The study was approved by Rennes University Hospital ethics committee and is registered at https://www.clinicaltrials.gov (NCT02286024). The patients suffered from MD and all of them were experiencing a major depressive episode according to DSM-5 criteria (Montgomery and Åsberg Depression Rating Scale (MADRS) (Montgomery and Åsberg, 1979) ≥ 15) when recruited (American Psychiatric Association, 2013). They were free of any other axis-I disorders (except for anxious comorbidities such as post-traumatic stress disorder, social phobia, generalized anxiety disorder or panic disorder) as determined by the Mini-International Neuropsychiatric Interview (Sheehan et al., 1998). Additionally, all participants with any severe chronic physical illness, diagnosed with neurodegenerative disorders (e.g. Parkinson's disease, Alzheimer's disease, Huntington's disease) or dementia (according to DSM-5 criteria) were excluded from this study. Other exclusion criteria involved potential safety contraindications for MRI (pacemakers, metal implants, pregnancy and lactation), addiction to alcohol and tobacco, or a history of significant head injury (history of head injury with loss of consciousness requiring hospitalization). No axis-II conditions were tested or excluded. For each patient, demographic data, comorbidities and medication, as well as clinical variables were collected (Table 1). A composite measure of medication load for each patient was assessed using a previously established method (Sackeim, 2001), taking into account the number of medication classes prescribed to the patients as well as each dosage.

In addition, 77 sex- and age-matched HC with no psychiatric disorders were recruited. The same MRI sequences were performed as for the patients.

## 2.2 Clinical assessment

All patients were assessed once at baseline and a subset of them at 6-month follow-up ($n = 69$) by a trained senior psychiatrist. The Clinical Global Impression-Improvement (CGI-I) scale was used to measure treatment response (Guy, 1976), considering the patient's history, social circumstances, symptoms and the impact of the illness on the patient's ability to function (Busner and Targum, 2007). The CGI-I item has been repeatedly reported as a suitable scale for everyday-clinic practice. It has been validated and well correlated with other standard scales such as MADRS clinician rating scale (Busner and Targum, 2007; Min et al., 2012; Mohebbi et al., 2018). A meta-analysis has shown it to be reliable (Spielmans and Mcfall, 2006) and other authors pointed out that it was a good predictor of the long-term outcome of depression (Simon, 2000). Based on previous studies (Leucht et al., 2013; Min et al., 2012), the I and NI groups were created by cutting off the CGI-I score at 2. Patients with a score of 1 or 2 were considered as I (i.e., very much improved or much improved), while those with a score of 3 to 7 were considered as NI, resulting in 36 I and 33 NI. The 6-month timescale was chosen because it corresponds to the median duration of a mood depressive episode (Have et al., 2017). In addition, other clinical scales were assessed: Clinical Global Impression-Severity (CGI-S) (Busner and Targum, 2007) and MADRS to measure depression severity, State Trait Anxiety Inventory A and B (STAI) (Spielberger, 1970) to measure state and trait anxiety, Snaith Hamilton Pleasure Scale (SHAPS) (Snaith et al., 1995) to measure anhedonia, and Apathy Evaluation Scale (AES) (Marin et al., 1991) to measure apathy.

The treatments received by the patients at baseline and at 6-month follow-up are described in the Supplementary Information (Table S3).

### 2.3 MRI acquisition

MRI was carried out on the subjects within five days after clinical assessment on a 3T whole body Siemens MR scanner (Magnetom Verio, Siemens Healthcare, Erlangen, Germany) with a 32-channel head coil. The 3D T1-weighted image was acquired using the following parameters: TR = 1.9 s, TE = 2.26 ms, flip angle = 9°, in-plane resolution = 2 × 2 mm$^2$, 256 × 256 mm$^2$ field of view (FOV), and thickness/gap = 1.0/0 mm.

DWI was acquired on 60 slices using an interleaved slice acquisition with the following parameters: slice thickness of 2 mm without gap, in-plane resolution = 1 × 1 mm$^2$, 256 × 256 mm$^2$ FOV, acquisition matrix 128 × 128, reconstruction matrix 128 × 128, TR/TE = 11,000/99 ms, flip angle 90°, pixel bandwidth 1698 Hz, imaging frequency 128 MHz, 30 diffusion directions, and b-value 1000 s/mm².

### 2.4 Data preprocessing

The diffusion-weighted images were preprocessed using the open source medical image processing toolbox Anima (https://anima.irisa.fr), including the following steps:

1. The DWI images were corrected for eddy currents by registering linearly then non-linearly each sub-volume of the EPI series to the first sub-volume, ensuring an opposite symmetric transformation (Hédouin et al., 2017).
2. Rigid realignment was performed to correct for subject motion.
3. EPI distortion correction of the images was performed using the reversed gradient method (Voss et al., 2006), coupled with a symmetric block-matching distortion correction method (Hédouin et al., 2017).
4. Denoising was handled using blockwise non-local means filtering (Coupe et al., 2008).
5. Skull stripping was performed using an atlas-registration-based method by transforming the structural image of each patient to the atlas image, using the linear and non-linear block-matching algorithms (Commowick et al., 2012a; Ourselin et al., 2000).
6. Diffusion tensor estimation was performed using a log-Euclidean estimation method (Fillard et al., 2007).
7. Whole-brain white-matter tractography was achieved using fiber assignment by continuous tracking (FACT) method (Mori et al., 1999).

The T1-weighted images were coregistered to the b = 0 images in the DWI space (Commowick et al., 2012b). Finally, the parcellation of the cerebral cortex of each participant into 80 regions of interest including 66 cortical and 14 subcortical regions was performed using FreeSurfer (Dale et al., 1999).

## 2.5 Brain network construction and graph metrics

Connectivity matrices were constructed using the parcellation mentioned above. The connections between two regions *i* and *j* were obtained by calculating the total number of fiber tracts connecting *i* and *j* and normalizing them by the mean volume of those two regions. Those connections were further thresholded to include only links found in at least 70% of participants, in order to suppress false positive fibers arising from tractography errors (Maier-Hein et al., 2017).

Graph metrics were then computed on undirected weighted matrices to examine changes in network structure (e.g., reflecting segregation and integration). The following global graph metrics were calculated: 1) characteristic path length, which is computed as the average number of edges in the shortest path between all pairs of nodes in the graph, 2) global clustering coefficient, which is the ratio of the number of triangles (set of three different nodes where each node is connected to the other two) to the number of connected triplets (set of three nodes where consecutive nodes are connected to each other), and 3) global efficiency, which reflects the capacity for network-wide communication, computed as the inverse of the average characteristic path length between all nodes. To describe each region individually, the following local graph metrics were also considered: 1) clustering coefficient, which is the probability that two neighbors of a given node are also connected to each other, 2) local efficiency, which is the inverse of the shortest average path length of all neighbors of a given node among themselves, 3) degree centrality, which is the number of edges connected to a node, 4) betweenness centrality, which is the fraction of all shortest paths that pass through a given node, 5) participation coefficient, which determines if the edges of a node tend to be clustered into one module of the network or link different other modules, computed as the ratio of the number of edges connecting a given node within its module to its degree, 6) node curvature, which is derived from the concept of graph curvature to analyze brain robustness (Farooq et al., 2019), and 7) strength, which is the sum of the weights of the adjacent edges of a given node.

## 2.6 Statistical analysis

A general linear model was applied to the edges as well as to the global and local metrics to remove the potential effects of age and sex covariates. Global and local properties of the structural brain networks were analyzed using the bctpy toolbox with Python (https://github.com/aestrivex/bctpy). The pipeline of the statistical analysis is summarized with each step detailed in Figure 1.

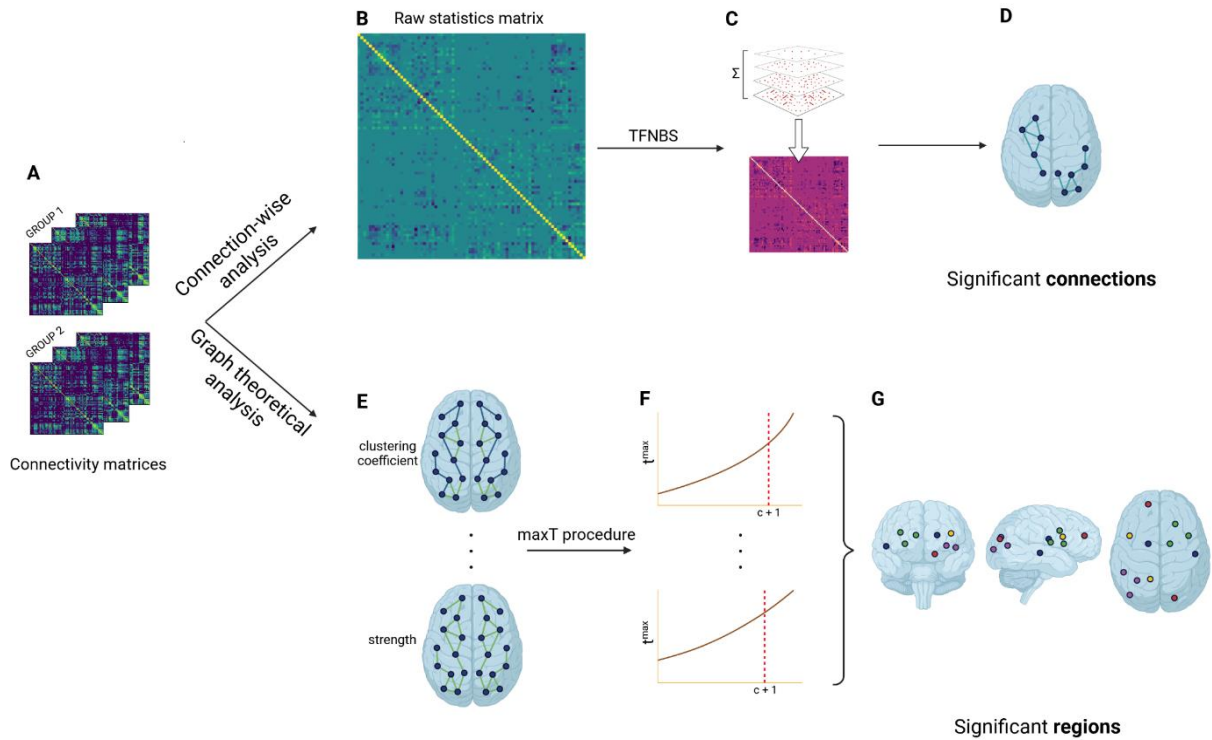

**Figure 1.** (**A**) Connectivity matrices of the groups studied, mood disorder patients ($n = 154$) vs. healthy controls ($n = 77$), or improved ($n = 36$) vs. not-improved ($n = 33$). (**B**) Raw statistics matrix obtained by performing *t*-tests on each edge between the groups. The matrix in (**C**) is obtained after performing the TFNBS algorithm: the raw statistics matrix is thresholded at a series of steps where connected components are identified at each step. The subsequent matrices across all steps are then summed to obtain the final TFNBS matrix. Statistical significance is established through permutation testing by building the null distribution of the maximum TFNBS scores across the links, and comparing them to the observed TFNBS matrix, resulting in FWE-corrected *p*-values. (**D**) Significant connections from each connected component. (**E**) Computation of global and local graph metrics (global: characteristic path length, global clustering coefficient, global efficiency, local: clustering coefficient, local efficiency, degree centrality, betweenness centrality, participation coefficient, node curvature, strength). (**F**) Correction of multiple comparisons across the brain regions by using the maxT procedure (100000 permutations). (**G**) Significant regions retrieved after the maxT procedure. Each color represents a different metric (arbitrarily, here).

*2.6.1 TFNBS*

The TFNBS combines the NBS and the threshold-free cluster enhancement algorithm for voxel-wise analysis adapted to graphs. A raw statistics matrix *M* is first computed by performing *t*-tests on each connection of the connectivity matrices between the two groups studied, before thresholding it at different steps *h* with intervals *dh*. Then, at each threshold *h*, we find the connected components and count the number of connections *e(h)* in each of them. Finally, the entries in *M* are replaced by their TFNBS score, producing the final TFNBS score matrix. The TFNBS score of edge *i, j* is calculated as follows:

$$\text{TFNBS}(i,j) = \int_0^{h(i,j)} e(h)^E h^H dh,$$

with *h(i,j)* the value of the considered entry *M(i,j)* at threshold *h*, *E* and *H* the extension and height empirical parameters, respectively. We defined the parameters *E* = 0.4 and *H* = 3.0 for the comparison between MD patients and HC, and *E* = 0.25 and *H* = 3.1 for assessing I and NI differences. The parameters have been chosen empirically. Grids are illustrated in Supplementary Figure S1 and S2 (see (Baggio et al., 2018; Vinokur et al., 2015) for a discussion about parameter choice). Then, we performed 5000 permutations to correct for multiple comparisons by building the permutation distribution of the maximum TFNBS scores across the edges and compared the observed TFNBS matrix with it. This resulted in *p*-values associated with each entry in the final TFNBS matrix.

*2.6.2 Topological metrics*

A *t*-test was applied for each global metric to examine disruptions in the topological organization of whole-brain networks (*p* < .05). The same test was performed on each region for the local metrics to identify the regions displaying significant differences between the groups. To correct for multiple comparisons, the maxT procedure using nonparametric permutation tests was considered (100000 permutations) (Westfall and Young, 1993). At each permutation *i* = 1, …, *N*, the labels were exchanged randomly between the two groups studied. Then, the maximal statistic over all regions $t_i^{max}$ was saved for permutation *i*, thus generating a permutation distribution of maximal statistics (Nichols and Holmes, 2002). Finally, a parameter $\alpha$ = .05 was chosen in order to obtain a critical threshold as the $c$ + 1 largest member of the distribution, with $c = \lfloor \alpha N \rfloor$. All statistics exceeding this threshold were considered to be significant.

## 3 Results

### 3.1 Demographic and clinical measures

The groups of HC and MD patients were comparable in terms of age (HC: 48.7 $\pm$ 12.6, MD: 50.8 $\pm$ 16.7; *t*-test *p* = 0.33) and sex (HC: 28M/49F, MD: 56M/98F; Chi-square *p* = 0.47), as well as the I and NI groups for age (*p* = .068) and sex (*p* = 0.44). The I and NI groups did not reveal any significant differences for the clinical scales tested at baseline, except for trait anxiety (Table 1). Indeed, the NI group suffered from higher baseline trait anxiety than the I group (*p* = .003). At baseline and at 6-month follow-up, there were no significant differences except for the clinical variables (see Supplemental Table S4). Socio-demographic data, medication load and clinical characteristics with the relevant statistical tests at the two time points for the I and NI groups are summarized in Table 2.

### 3.2 Whole-brain mapping of connectivity deficits

Between MD patients and HC, the TFNBS revealed 15 significant connections including the media orbitofrontal, precuneus, middle temporal, left and right superior parietal, the thalamus of both hemispheres, the hippocampus, the caudate and the right putamen, as displayed in Figure 2. All these connections were found to exhibit lower connectivity among MD participants except between the right thalamus and the right caudate. A single edge including the left inferior temporal and the left insula was found between the I and NI groups, showing increased connectivity among the I group (Figure 2).

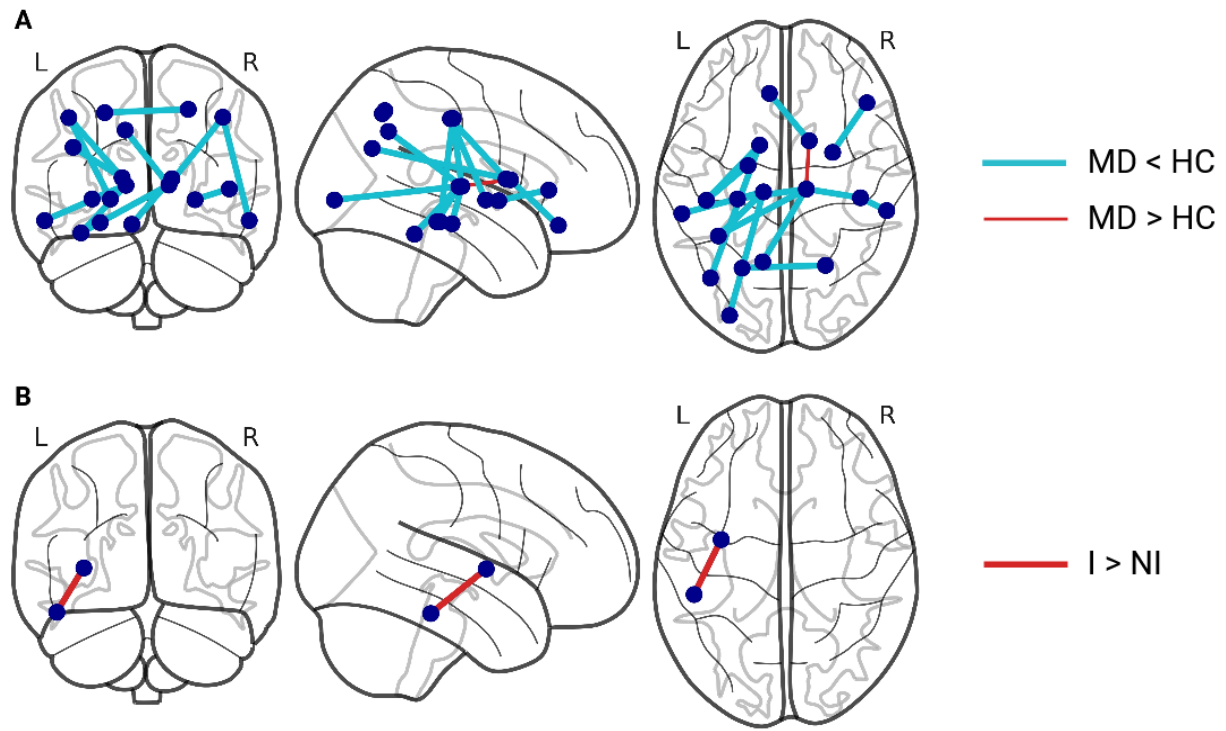

**Figure 2.** Significant connections characterizing mood disorder and treatment-resistance depression using the threshold-free network-based statistics. **(A)** Significant connections between depressed patients versus healthy controls. All the connections display significantly lower connectivity (blue) except the one between the right thalamus and the right caudate (red). **(B)** Significant connection between improved and not-improved, linking the left inferior temporal and the left insula. HC, healthy controls; L, left; MD, mood disorder; NI, not-improved; R, right; I, improved.

### 3.3 Whole-brain differences in topological metrics

Compared with HC, MD patients showed significantly higher global characteristic path length ($p < 10^{-5}$) and global clustering coefficient ($p = .008$), as well as lower global efficiency ($p < 10^{-6}$). We further investigated local metrics to identify key regions that might lie behind the global network alterations. A higher capacity of possible information transfer was found in MD patients as measured with local efficiency in regions involved in emotional processing such as the right lingual gyrus, insula in both hemispheres and the right amygdala. Local clustering coefficient also demonstrated such variations in these regions. Metrics related to network centrality displayed few significant different regions, namely the caudal middle frontal and rostral anterior cingulate (RAC) for degree centrality, and the thalamus as well as the globus pallidus for betweenness centrality. Overall, the regions identified by the topological metrics were similar (Figure 3). For details on the significant regions identified, see Supplemental Table S1. Between the I and NI groups, none of the topological metrics showed significant differences after correcting for multiple comparisons. At an uncorrected threshold $p = .05$, the measures exhibited fewer significant regions than between the MD and HC groups, except for those belonging to network centrality (degree centrality, betweenness centrality, participation coefficient). Those regions were mainly included in the frontal, parietal lobes and the DMN with the left precuneus, right middle temporal lobe and the right RAC (Figure 4; Table 3). The full list of significant regions is presented in Supplemental Table S2.

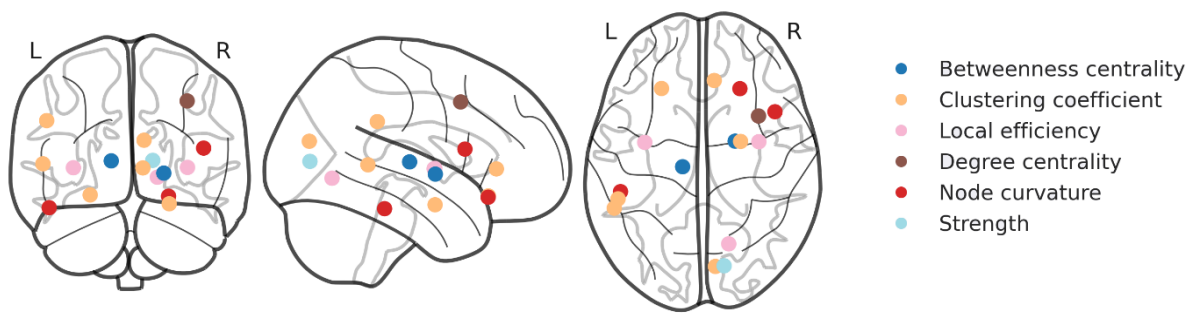

**Figure 3.** Regions exhibiting significant differences between the mood disorder and healthy control groups according to the results obtained with the maxT procedure on the topological metrics.

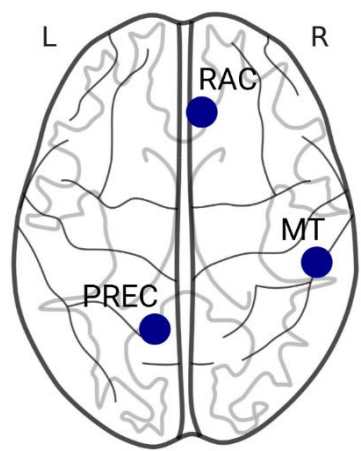

**Figure 4.** Significant regions that characterize differences between improved and not-improved using *t*-tests on graph metrics at an uncorrected threshold ($p < .05$). All of the regions belong to the default mode network. L, left; MT, middle temporal; PREC, precuneus; R, right; RAC, rostral anterior cingulate.

## 4 Discussion

In this study, we identified alterations in the thalamus as well as regions belonging to the basal ganglia, namely the caudate and putamen, which are predominately involved in the limbic-cortical-striatal-pallidal-thalamic network in MD patients compared to HC. This is in agreement with previous studies that found disruptions in these regions by means of widespread volume reductions in MDD patients, specifically the bilateral thalamus (Arnone et al., 2016), left putamen and right caudate (Korgaonkar et al., 2014; Lu et al., 2017). There is growing evidence that dysfunctions in these circuits may regulate the disturbances in autonomic regulation and neuroendocrine responses associated with mood disorders (Herane-Vives et al., 2018; Peng et al., 2013). The identification of such predictors is clinically relevant, as detecting neuroendocrine dysregulation could help clinicians tailor treatments used for depressed patients. Besides, another work highlighted the role of the basal ganglia in cognitive and emotional behaviors, showing that alterations in these networks could lead to depressive disorder (Haber, 2022). In addition, the insula and amygdala were found to display important differences between MD patients and HC, as well as I and NI. These are the core regions belonging to the fronto-limbic network and are involved in emotion regulation. Indeed, they are part of the salience network (SN), a large-scale brain network known to contribute to many functions such as self-awareness through the integration of sensory, emotional, and cognitive information (Menon, 2015). Hamilton et al. further highlighted the crucial role of this network, suggesting that the occurrence of repetitive, preservative, negative thinking and biases in attention to negative events may underlie aberrant SN response and connectivity in depression (Hamilton et al., 2013).

The present study investigated the topological organization of structural brain networks in MD patients compared to HC. The results revealed that MD patients exhibited increases in both characteristic path length and global clustering coefficient. These metrics are parameters of the small-worldness that reflect the intrinsic principles in a network: integration and segregation (Sporns, 2011). Small-world networks are characterized by high clustering and short path length. Here, the changes in the global metrics indicated significant increases in small-world features in the brain networks of depressed patients, suggesting a less integrated and more segregated organization in the whole brain networks. Moreover, the local clustering coefficient in MD patients displayed higher values in important regions such as the lingual gyrus, insula and amygdala compared with HC. Likewise, the lower global efficiency observed in the MD group is relevant to a disrupted small-world organization. These changes could be associated with a reorganization of the structural brain networks centered around the limbic system or the SN with longer distance connections, suggesting that information transfer is predominant across those networks in depression. This is in agreement with a previous study that identified functional and structural alterations in limbic regions in youth depression, especially in the amygdala (Redlich et al., 2018). Another study associated a stronger clustering coefficient of the right amygdala with higher levels of depression symptoms levels and duration of depressive episodes among MDD patients (Jacob et al., 2022). These findings might reflect the attenuation of positive information contrasted with intensification of affective information, according to cognitive theories of depression (Beck, 2008; Stuhrmann et al., 2013). Such modifications in small-world parameters and network efficiency are also found in other brain diseases, such as epilepsy (van Diessen et al., 2014) or Parkinson's disease (Pereira et al., 2015), describing widespread changes affecting the whole network and a reorganization of the networks' hubs in diseases. We also observed increasing local metrics in MD patients, as measured by local efficiency and betweenness centrality in the right amygdala and the left thalamus, respectively. This is in line with a previous study suggesting that both regions are involved in emotional perception (Ye et al., 2015). Our findings additionally revealed a decrease in degree centrality in the left cuneus and an increase in local efficiency in the right

RAC, which have not yet been reported by other DWI studies. Other results reported an increased degree centrality in these regions but in functional connectivity, which may encourage us to further study the relationship between structural and functional connectivity (Ye et al., 2015; Zhang et al., 2011).

Abnormal changes in the DMN connectivity is well established in TRD (Li et al., 2013). Within the network, connectivity changes are deemed as a plausible explanation of brain responses associated with cognitive demands and processing of emotions in depression (Wise et al., 2017). As such, the network has been hypothesized as a predictor to treatment response due to its role in the generation of self-referential processing, negative rumination, emotion regulation, and depressive symptoms (Long et al., 2020; Wise et al., 2017). Compared to the NI group, we found a reduction in local clustering coefficient, efficiency and strength of the RAC in the I group using graph theoretical analysis. The RAC is a key hub of the DMN, for which previous studies indicated hyperactivation or hyperconnectivity related to depressive symptom improvements in functional connectivity (Gerlach et al., 2022; Jamieson et al., 2022). A study examining electroencephalography (EEG)-based connectivity in theta and beta bands reported that the RAC and the insula were predictors of depression remission (Whitton et al., 2019). Previous studies also suggested that the functional activity of the RAC is a therapeutic target of choice in noninvasive brain stimulation (Weigand et al., 2018) and invasive deep brain stimulation (Mayberg et al., 2005) for TRD. Indeed, numerous studies have investigated the effects of treatment on functional connectivity in the DMN and its relationship with clinical improvement (de Kwaasteniet et al., 2015; Pang et al., 2022). Although these works did not analyze graph metrics, their results might still indicate important regions that characterize response to antidepressants (Chin Fatt et al., 2020; Lui et al., 2011). More specifically, a previous study reported a correlation between increased within-network functional connectivity of lateral DMN regions including the precuneus, superior and middle temporal cortex, and improvement in depression (Kilpatrick et al., 2022). A prior study focusing on late-life depression reported increased DMN functional connectivity for posterior DMN regions such as the precuneus following antidepressant treatment (Andreescu et al., 2013). Another similar study showed a correlation between increased DMN functional connectivity in lateral DMN regions like the middle temporal gyrus and treatment response to antidepressants (Karim et al., 2017).

To the best of our knowledge, our study is the first to identify the DMN as a key structural hub associated with treatment response over a 6-month follow-up. The present work revealed significant differences in the precuneus, the middle temporal lobe and the RAC, that are part of the DMN. In addition to the literature on functional connectivity patterns of DMN in TRD, our results suggest that white matter integrity might also be a determinant of treatment response as well. Few studies have investigated the relationship between structural and functional abnormalities associated with depression. One recent meta-analysis has pointed out a negative relationship as illustrated by reduced gray or white matter in the DMN associated to hyperconnectivity within the DMN in depressed patients as compared to HC (Scheepens et al., 2020). With our study, we can hypothesize that treatment response could be associated with both DMN structural and functional connectivity as well. Furthermore, some authors have suggested that functional brain activity (as measured by EEG) may depend on underlying structural integrity (Babaeeghazvini et al., 2021). Based on our work and the literature on DMN functional connectivity in TRD, we could then hypothesize that network structural integrity could be a necessary condition for functional response and, by consequence, clinical improvement. Multimodal studies are warranted to test this dual condition.

There may be some possible limitations in this study. The population was not perfectly equally distributed between women and men but it remains representative of depression sex ratio as illustrated by a higher prevalence in women (Kessler, 2003). It can be noted that our sample presented some aspects of heterogeneity in terms of clinical profile (bipolar depression, anxiety disorders), which may have limited our statistical power. The presence of comorbidities and medication is also important in this regard. However, our sample is representative of the TRD population (Dudek et al., 2010; Kautzky et al., 2019), which strengthens the generalizability of our findings. This study should be replicated in a larger population of I and NI to assess whether such analysis could provide clearer information on brain changes between these two groups. Another limitation is its cross-sectional design where the brain imaging was performed at a single time point, which precludes causal inferences or capturing progressive changes over time. Thus, it would be interesting to corroborate our findings in longitudinal studies. In the future, it would be helpful to combine multimodal imaging data to study the interaction between functional and structural connectivity as more networks contributing to treatment response such as the DMN might be identified.

**Acknowledgments**

MRI data acquisition was supported by the Neurinfo MRI research facility from the University of Rennes I. Neurinfo is granted by the European Union (FEDER), the French State, the Brittany Council, Rennes Metropole, Inria, Inserm and the University Hospital of Rennes. This work has been funded by Institut des Neurosciences Cliniques de Rennes (INCR). The authors thank Mr Stéphane Brousse and Mr Jacques Soulabaille for their involvement in the conduct of the study.

**Author Contributions Statement**

**Sébastien Dam**: Conceptualization; Formal analysis; Methodology; Software; Visualization; Writing - original draft. **Jean-Marie Batail**: Data curation; Funding acquisition; Investigation; Resources; Validation; Writing - review & editing. **Gabriel H. Robert**: Data curation; Investigation; Resources; Writing - review & editing. **Dominique Drapier**: Data curation; Funding acquisition; Investigation; Resources. **Pierre Maurel**: Conceptualization; Methodology; Project administration; Resources; Supervision; Validation; Writing - review & editing. **Julie Coloigner**: Conceptualization; Methodology; Project administration; Resources; Supervision; Validation; Writing - review & editing.

**Author Disclosure Statement**

Declarations of interest: none

**Funding Statement**

This work has been funded by Institut des Neurosciences Cliniques de Rennes (INCR).

**Table 1.** Intergroup comparison of demographic and clinical characteristics of the mood disorder, improved and not-improved groups at baseline.

|  | MD group (n = 154) | | | I group (n = 36) | | | NI group (n = 33) | | | |
|---|---|---|---|---|---|---|---|---|---|---|
|  | Mean / n | SD / % | n | Mean / n | SD / % | n | Mean / n | SD / % | n | p-value |
| *Sociodemographic variables* | | | | | | | | | | |
| Age (years) | 50.8 | 16.7 | 154 | 56.2 | 16.6 | 36 | 49.2 | 14.6 | 33 | 0.068b |
| Sex (M/F) | 56/98 | | 154 | 10/26 | | 36 | 10/23 | | 33 | 0.443c |
| *Comorbidities* | | | | | | | | | | |
| Bipolar disorder: | 85 | 58.2% | 146 | 13 | 36.1% | 36 | 14 | 42.4% | 33 | 0.443c |
| type 1 | 12 | 8.2% | 146 | 1 | 2.8% | 36 | 3 | 9.1% | 33 | 0.343d |
| type 2 | 29 | 19.9% | 146 | 4 | 11.1% | 36 | 9 | 27.3% | 33 | 0.126d |
| type 3 | 20 | 13.7% | 146 | 8 | 22.2% | 36 | 2 | 6.1% | 33 | 0.087d |
| Panic disorder | 45 | 30% | 150 | 13 | 36.1% | 36 | 13 | 39.4% | 33 | 0.443c |
| Generalized anxiety disorder | 64 | 42.7% | 150 | 19 | 52.8% | 36 | 15 | 45.5% | 33 | 0.443c |
| Social phobia | 18 | 12.1% | 149 | 4 | 11.1% | 36 | 3 | 9.1% | 33 | 0.702d |
| PTSD | 14 | 9.3% | 150 | 2 | 5.6% | 36 | 4 | 12.1% | 33 | 0.141d |
| *Medication* | | | | | | | | | | |
| Medication load | 3.0 | 1.3 | 147 | 3.2 | 1.2 | 36 | 3.1 | 1.3 | 33 | 0.639b |
| Antidepressant | 71.4% | | 147 | 28 | 77.8% | 36 | 20 | 60.6% | 33 | 0.209c |
| Antipsychotic | 33.2% | | 147 | 7 | 19.4% | 36 | 12 | 36.4% | 33 | 0.782c |
| Benzodiazepine | 66.2% | | 146 | 28 | 77.8% | 36 | 23 | 69.7% | 33 | 0.292c |
| *Clinical variables* | | | | | | | | | | |
| Duration of illness (years) | 16.5 | 14.5 | 148 | 21.6 | 18.5 | 35 | 13.5 | 10.5 | 33 | 0.132b |
| Number of episodes | 4.99 | 4.7 | 147 | 4.8 | 3.8 | 34 | 5.9 | 6.3 | 33 | 0.791b |
| MADRS | 26.9 | 6.1 | 151 | 27.0 | 6.0 | 36 | 27.2 | 5.2 | 33 | 0.877a |
| STAI-YA | 58.2 | 12.2 | 149 | 58.5 | 11.4 | 35 | 57.7 | 11.8 | 32 | 0.798a |
| STAI-YB | 58.8 | 12.7 | 150 | 58.8 | 10.6 | 35 | 65.6 | 9.6 | 33 | 0.008b |
| SHAPS | 5.6 | 3.9 | 149 | 5.6 | 3.8 | 35 | 5.5 | 3.7 | 33 | 0.975b |
| AES | 42.1 | 9.5 | 150 | 43.7 | 9.3 | 36 | 44.0 | 10.7 | 33 | 0.995b |

a:t-test; b: Mann-Whitney U test; c: Chi2 test; d: Fisher test; AES, apathy evaluation score; MADRS, Montgomery and Åsberg depression rating scale; SD, standard deviation; SHAPS, Snaith Hamilton pleasure scale; STAI, state trait anxiety inventory.

**Table 2.** Improved and not-improved description at baseline and at 6-month follow-up.

| Variables (*n = 69*) | M0 | | | | M6 | | | | |
|---|---|---|---|---|---|---|---|---|---|
| | Mean / *n* | SD / % | *n* | Range | Mean / *n* | SD / % | *n* | Range | *p*-value |
| *Sociodemographic variables* | | | | | | | | | |
| Age (years) | 52.9 | 15.9 | 69 | 21-87 | | | | | - |
| Sex (M/F) | 20/49 | | 69 | | 20/49 | | 69 | | - |
| *Comorbidities* | | | | | | | | | |
| Bipolar disorder: | 27 | 40.3% | 67 | | 20 | 38.5% | 52 | | 0.625a |
| type 1 | 4 | 5.9% | 67 | | 2 | 3.8% | 52 | | 1a |
| type 2 | 13 | 19.4% | 67 | | 9 | 17.3% | 52 | | 1a |
| type 3 | 10 | 14.9% | 67 | | 9 | 17.3% | 52 | | 1a |
| Panic disorder | 26 | 37.7% | 69 | | 23 | 41.1% | 56 | | 0.727a |
| Generalized anxiety disorder | 34 | 49.3% | 69 | | 19 | 33.9% | 56 | | 0.238a |
| Social phobia | 7 | 10.3% | 68 | | 8 | 14.3% | 56 | | 1a |
| PTSD | 6 | 8.7% | 69 | | 5 | 8.9% | 56 | | 1a |
| *Medication* | | | | | | | | | |
| Antidepressant | 48 | 72.7% | 66 | | 45 | 65.2% | 69 | | 0.508a |
| Antipsychotic | 19 | 28.8% | 66 | | 25 | 36.2% | 69 | | 0.388a |
| Benzodiazepine | 51 | 78.5% | 65 | | 33 | 47.8% | 69 | | 0.057a |
| *Clinical variables* | | | | | | | | | |
| Duration of illness (years) | 17.6 | 15.7 | 68 | 0-60 | 18.0 | 15.8 | 49 | 0.75-60 | - |
| Number of episodes | 5.3 | 5.3 | 67 | 0-30 | 5.2 | 5.3 | 53 | 1-31 | 0.957b |
| MADRS | 27.1 | 5.7 | 69 | 16-43 | 15.2 | 10.4 | 68 | 0-40 | < 10$^{-5}$b |
| STAI-YA | 58.1 | 11.6 | 67 | 28-78 | 46.2 | 17.7 | 66 | 20-78 | < 10$^{-5}$b |
| STAI-YB | 62.1 | 10.7 | 68 | 37-79 | 55.2 | 14.8 | 66 | 24-79 | < 10$^{-5}$b |
| SHAPS | 5.5 | 3.8 | 68 | 0-14 | 3.7 | 4.0 | 65 | 0-14 | < 10$^{-5}$b |
| AES | 43.9 | 10.0 | 69 | 24-69 | 39.4 | 12.6 | 64 | 18-68 | 0.002b |

*a: McNemar test; b: Wilcoxon test. AES, apathy evaluation score; CGI-S, clinical global impression-severity; MADRS, Montgomery and Åsberg depression rating scale; SD, standard deviation; SHAPS, Snaith Hamilton pleasure scale; STAI, state trait anxiety inventory.*

**Table 3.** The topological measures' change between the improved and not-improved groups

| ROIs | CC | LE | BC | NC | S |
|---|---|---|---|---|---|
| PREC | NS | ↗ | NS | ↗ | ↗ |
| MT | ↘ | ↘ | ↗ | ↘ | ↘ |
| RAC | ↘ | ↘ | NS | NS | ↘ |

*Up and down arrows indicate significant increased and decreased of the mean topological metric in improved group compared to not-improved group obtained with two-sample t-test (uncorrected p < .05). The acronym NS means no significant differences were found.*
*CC, clustering coefficient; LE, local efficiency; BC, betweenness centrality; NC, node curvature; S, strength; PREC, precuneus; MT, middle temporal; RAC, rostral anterior cingulate; ROIs, regions of interest.*